\documentclass[journal,12pt,onecolumn,draftcls]{IEEEtran}

\usepackage{cite}
\usepackage{amssymb}
\usepackage{graphicx,epstopdf}
\usepackage{amsmath}

\usepackage{amssymb}

\usepackage{amsmath}
\begin{document}
\title{Performance Analysis of Low-Density Parity-Check Codes over 2D Interference Channels via Density Evolution}

\author{Jun~Yao,
        ~Kah~Chan~Teh,~\IEEEmembership{Senior Member,~IEEE},
        and~Kwok~Hung~Li,~\IEEEmembership{Senior Member,~IEEE}}% <-this % stops a space
\maketitle

\begin{abstract}
The theoretical analysis of detection and decoding of low-density parity-check (LDPC) codes transmitted over channels with two-dimensional (2D) interference and additive white Gaussian noise (AWGN) is provided in this paper. The detection and decoding system adopts the joint iterative detection and decoding scheme (JIDDS) in which the log-domain sum-product algorithm is adopted to decode the LDPC codes. The graph representations of the JIDDS are explained. Using the graph representations, we prove that the message-flow neighborhood of the detection and decoding system will be tree-like for a sufficiently long code length. We further confirm that the performance of the JIDDS will concentrate around the performance in which message-flow neighborhood is tree-like. Based on the tree-like message-flow neighborhood, we employ a modified density evolution algorithm to track the message densities during the iterations. A threshold is calculated using the density evolution algorithm which can be considered as the theoretical performance limit of the system. Simulation results demonstrate that the modified density evolution is effective in analyzing the performance of 2D interference systems.
%\boldmath
\end{abstract}

\begin{IEEEkeywords}
LDPC codes, 2D ISI, density evolution, sum-product algorithm, factor graph, turbo equalization.
\end{IEEEkeywords}

% \fi
\IEEEpeerreviewmaketitle

\section{Introduction}
\IEEEPARstart{D}{uring} the past decades, much research has been done in searching for good error-correcting codes (ECCs). Turbo codes, proposed by Berrou et al. in 1993 \cite{ref1}, exhibit superior capabilities in error correcting and approach the Shannon limit by only 0.7 dB. Another class of ECCs, called low-density parity-check (LDPC) codes, was first proposed by Gallager in \cite{ref2}. Inspired by the success of turbo codes, the potentials of LDPC codes were re-examined in the mid-1990's with the work of MacKay, Luby, and others \cite{ref3}-\cite{ref5}. Comparing to Turbo codes, the LDPC codes have lower decoding complexity and shorter latency. But these regular LDPC codes perform about 0.5 dB worse than turbo codes. The construction of irregular LDPC codes was proposed by Luby et al. in \cite{ref6}, \cite{ref7}, which can be used to further improve the performance of LDPC codes. In \cite{ref8}, Richardson and Urbanke proposed a method, called density evolution, to analyze the asymptotic performance of LDPC codes over various memoryless channels. By using this method to optimize the degree distributions of LDPC codes, it was shown in \cite{ref9} that the LDPC codes can approach the Shannon limit by 0.0045 dB. In \cite{ref10}, a Gaussian approximation method was proposed which can simplify the analysis of the LDPC decoding algorithm.

The application of LDPC codes over one-dimensional (1D) inter-symbol interference (ISI) channel was investigated in \cite{ref11}-\cite{ref13}. In the joint detection and decoding scheme, the ISI channel is considered as the inner code and the LDPC codes serve as the outer code. The detector and decoder exchange extrinsic information and work iteratively in the same way as inner and outer decoders of turbo codes. Thus, this scheme is called the turbo equalization. In \cite{ref14}, the density evolution technique was adopted to analyze the message error probability in LDPC coded ISI channels. Based on \cite{ref14}, the authors modified the result to account for symbol errors in \cite{ref15}. In \cite{ref16}, the thresholds and scheduling of LDPC codes over ISI channels were investigated using the method of Gaussian approximation.

As the conventional magnetic storage systems are expected to encounter their storage density limits soon, two-dimensional (2D) interference arises with the detection of next generation ultra-high-density data storage systems, such as the patterned media storage (PMS) \cite{ref17,YaoWang2D} and two-dimensional magnetic recording (TDMR) \cite{ref18,JYAOTDMR} and heat-assisted magnetic recording (HAMR) \cite{ref19}. In these systems, the separation between adjacent tracks is so small that inter-track interference (ITI) cannot be ignored. Along with the conventional ISI, the system suffers from 2D interference which degrades the system performance significantly. Due to the lack of graph-based descriptions of 2D interference channels, there is no simple optimal detection algorithm for 2D interference channels, in contrast to the Viterbi algorithm \cite{viterbi} and the Bahl-Cocke-Jelinek-Raviv (BCJR) algorithm \cite{BCJR} for 1D ISI channels. It has been proved that maximum-likelihood sequence detection (MLSD) for 2D interference channels is NP-complete \cite{NP_complete}, which is too complex to be implemented. The performance bound of MLSD for 2D interference channels has been analyzed in \cite{JYAO}. Various suboptimal equalization and detection techniques have been proposed to mitigate the effect of 2D interference in \cite{NSingla}-\cite{JYAO4}. In \cite{ref21}-\cite{ref23}, different equalization techniques have been proposed to mitigate the effect of ITI. Several iterative detection algorithms have been proposed in \cite{ref24}-\cite{JYAO2}. Specifically, in \cite{IRCSDF1} and \cite{IRCSDF2}, the authors proposed an iterative row-column soft decision feedback algorithm (IRCSDFA) which is a concatenation of two soft-input/soft-output (SISO) detectors. In \cite{MCMC}, a Markov chain Monte Carlo (MCMC) based 2D detection algorithm was proposed. In this work, the \textit{a posteriori} bit probability is converted to the expected value of a function which is calculated using the Monte Carlo techniques. It has been shown that the MCMC-based algorithm outperforms the IRCSDFA that has been reported to have better performance than other existing detection algorithms. In \cite{JYAO2,JYAO3,JYAO4}, a joint iterative detection and decoding scheme (JIDDS) was proposed, which was also a concatenation of two constituent detectors. However, unlike the IRCSDFA in which the two detectors are concatenated in parallel and exchange weighted soft information, in JIDDS the two detectors are serially concatenated and the soft information exchanged between the two detectors is not weighted. It has been shown that the JIDDS outperforms the MCMC-based algorithm.

For 2D interference channels, we are interested in two questions: 1). Do LDPC codes also exhibit a threshold phenomenon over 2D interference channels? 2). What is the best performance so far that the LDPC coded 2D interference system can achieve? These two questions have not been answered in the literature, and they are addressed in this paper. Although the density evolution algorithm has been adopted to calculate the noise thresholds in \cite{NSingla}, the authors did not prove or even agree to the concentration results. They claimed that the concentration results do not hold due to the short cycles in the channel graph of their low-complexity joint equalization and decoding scheme (LCJEDS). At the same time, the authors showed in \cite{NSingla} that the simulation results for LCJEDS respect the thresholds calculated through the density evolution algorithm. In this paper, the JIDDS is adopted in the detection of LDPC coded 2D interference systems. Via the graph representation of the detection algorithm, we will prove a concentration result that as the code length tends to infinity, the average performance of the system is closely concentrated around the expected performance which exhibits a threshold phenomenon and it can be determined based on the tree-like message-flow neighborhood. By using a modified density evolution algorithm, a threshold is then calculated and can be considered as the supremum of performance for error-free transmission. When the codeword length is long enough, error-free performance can be achieved if and only if the parameter of the channel is below the calculated threshold. Simulation results verify that the thresholds can accurately predict the performance of the JIDDS for large block lengths. The rest of the paper is organized as follows. In Section II, we introduce the LDPC coded 2D interference system. In Section III, we briefly describe the algorithm of the JIDDS. In Section IV, we first demonstrate the graph representation of the detection algorithm. After that, two concentration statements are shown and the density evolution and threshold calculation are explained. Numerical results and discussions are presented in Section V. Finally, we conclude the paper in Section VI.

\section{System Model}
A discrete-time 2D ISI channel can be represented by a channel response matrix with a size of $M_{h}\times N_{h}$ given by
\begin{eqnarray}
\label{1}
\textbf{H}=[\textbf{h}_0,\textbf{h}_1,...,\textbf{h}_{N_{h}-1}]=\begin{bmatrix}
h(0,0) &h(0,1)  &...&h(0,N_{h}-1) \\
h(1,0) &h(1,1)  &...&h(1,N_{h}-1) \\
.. &  ..& &..\\
 h(M_{h}-1,0)&  h(M_{h}-1,1) &...&h(M_{h}-1,N_{h}-1)
\end{bmatrix}
\end{eqnarray}
where each $\textbf{h}_i$ is an $M_{h}$-tuple of real numbers, and $M_{h}$ and $N_{h}$ are the numbers of signals sensed by the read head in the cross-track and down-track directions, respectively.

\begin{figure}[!t]
\centering
\includegraphics[width=5in]{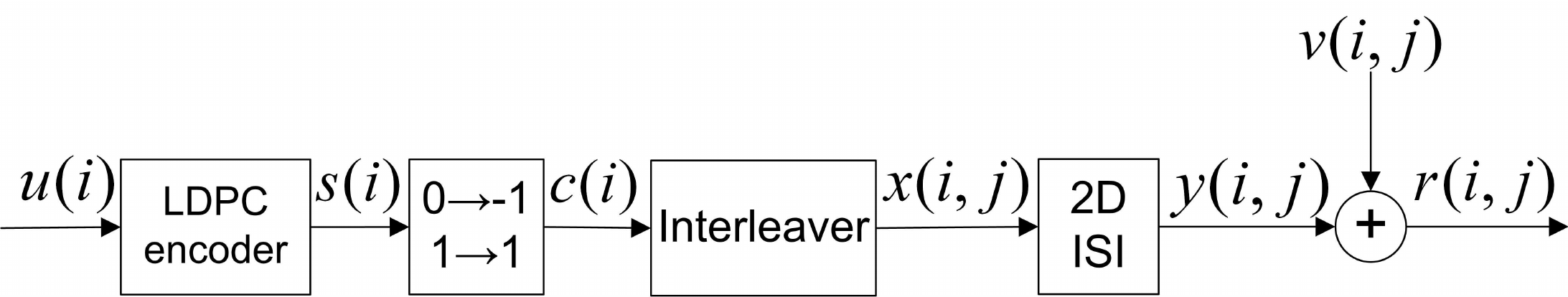}
% where an .eps filename suffix will be assumed under latex,
% and a .pdf suffix will be assumed for pdflatex; or what has been declared
% via \DeclareGraphicsExtensions.
\caption{Read channel model of the LDPC coded 2D interference system.}
\label{figure1}
\end{figure}

The channel model of a 2D recording system is illustrated in Fig. \ref{figure1}. An array of binary bits $\textbf{u}=\{ u(i)\in \{ 0, 1 \}: i=1,2,...,K\}$ is first passed to the LDPC encoder. Let the parity-check matrix of the LDPC code be denoted by $\textbf{H}_c$, and the corresponding generator matrix be denoted by $\textbf{G}_c$ (with the property $\textbf{H}_c \cdot \textbf{G}_c = \textbf{0}$). Here we consider a coset code of LDPC code which is specified by a parity-check matrix $\textbf{H}_c $ and a $1\times N$ coset-defining vector $\textbf{b}$, given by
\begin{eqnarray}
\label{coset1}
\textbf{s}=[s(1),s(2),...,s(N)]=[ \textbf{G}_c\cdot\textbf{u}]^T\oplus \textbf{b}
\end{eqnarray}
where $\oplus$ represents binary addition. The codeword $\textbf{s}$ satisfies
\begin{eqnarray}
\label{coset2}
\textbf{H}_c\cdot \textbf{s}^T=\textbf{d}^T=[d(1),d(2),...,d(N-K)]^T=\textbf{H}_c\cdot \textbf{b}^T.
\end{eqnarray}
The code is linear if and only if $\textbf{d}=\textbf{0}$; otherwise, this code is a coset code of an LDPC code. The node degree distribution of the code is specified by two degree polynomials \cite{ref7}
\begin{eqnarray}
\label{coset3}
\lambda (x)=\sum_{i=1}^{d_v^{max}} \lambda _ix^{i-1}~\textup{and}~\rho (x)=\sum_{i=1}^{d_c^{max}} \rho _ix^{i-1}
\end{eqnarray}
where  $d_v^{max}$ and $d_c^{max}$ are the maximal variable-node and check-node degrees, respectively. The parameters $\lambda _i$ and $\rho _i$ represent the fraction of edges with variable-node degree $i$ and check-node degree $i$, respectively. A regular LDPC code is a code for which $\lambda _{d_v^{max}}=\rho _{d_c^{max}}=1$. We denote $N_e$ as the total number of variable-to-check edges in the graph of LDPC codes.

The 1D encoded array $\textbf{s}=\{ s(i)\in \{ 0, 1 \}: i=1,2,...,N\}$ is first bipolar modulated to $\textbf{c}$ with $c(i)=2s(i)-1$ and then distributed on a 2D array $ \{ x(j,k)\in \{ -1, 1 \}: j=1,2,...,N_{r}, k=1,2,...,N_{c}\}$ with $N=N_{r}\times N_{c}$ according to a 1D sequence to 2D array mapping scheme. This device is called the interleaver. The values $N_{c}$ and $N_{r}$ represent the number of columns and rows of the 2D array, respectively. The deinterleaver converts a 2D array back to 1D sequence. Due to the one-to-one correspondence between the variables $c(i)$ and $s(i)$, the soft information of these two variables are equivalent and will be used interchangeably. We assume the 2D array $\{ x(j,k): j=1,2,...,N_{r}, k=1,2,...,N_{c}\}$ is surrounded by $-1$s, i.e., $x(j,k)=-1$ for $ \{ j\leq 0, j\geq {N_{r}+1} \}$ or $ \{ k\leq 0, k\geq {N_{c}+1} \}$. The desired signal is interfered by $N_h-1$ symbols in the down-track direction and $M_h-1$ tracks in the cross-track direction, which can be expressed as
\begin{eqnarray}
\label{2}
y(i,j)&=&h(i,j)\ast x(i,j)\nonumber\\
&=&\sum_{m=0}^{M_{h}-1}\sum_{n=0}^{N_{h}-1}h(m,n)x(i-m,j-n)
\end{eqnarray}
where the 2D interference is generated by a 2D convolution of the channel response matrix with the corresponding interference in both along-track and cross-track directions. If we denote the cross-track direction $M_{h}$-tuple coded bits at the $i$-th track time instant $j$ as symbol $\textbf{x}_{i,j}=[x(i-M_{h}+1,j),x(i-M_{h}+2,j),...,x(i,j)]$, (\ref{2}) can be rewritten as
\begin{eqnarray}
\label{3}
y(i,j)=\sum_{n=0}^{N_{h}-1}\textbf{x}_{i,j-N_{h}+1+n}\cdot \textbf{h}_{n}.
\end{eqnarray}
The received signal is therefore equal to the 2D interfered signal plus additive white Gaussian noise (AWGN), given by
\begin{eqnarray}
\label{4}
r(i,j)=y(i,j)+v(i,j)
\end{eqnarray}
where $v(i,j)$ is AWGN with zero mean and variance $\sigma^2$. Thus, the signal-to-noise ratio (SNR) is defined as
\begin{eqnarray}
\label{5}
\textup{SNR}&=&10\cdot \log_{10}\left ( \frac{E_b}{N_0} \right )\nonumber\\
&=&10 \log_{10}\left ( \frac{\sum_{i=0}^{M_h-1}\sum_{j=0}^{N_h-1}\left ( h(i,j) \right )^2}{2 R \sigma ^{2}} \right ) \textup{dB}
\end{eqnarray}
where $E_b$ represents the energy per bit, $N_0$ is the power spectral density level of the AWGN, and $R=K/N$ is the overall code rate.

\section{Joint Iterative Channel Detection and LDPC Code Decoding}
\begin{figure}[!t]
\centering
\includegraphics[width=4in]{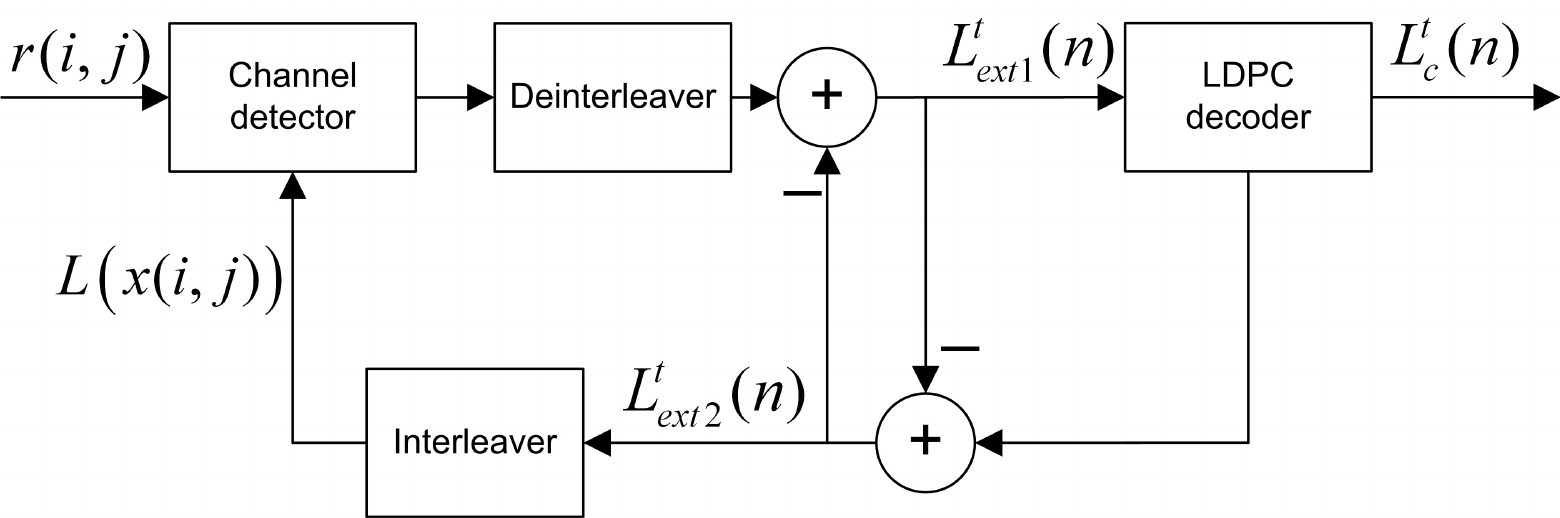}
% where an .eps filename suffix will be assumed under latex,
% and a .pdf suffix will be assumed for pdflatex; or what has been declared
% via \DeclareGraphicsExtensions.
\caption{System diagram of detector and decoder.}
\label{figure2}
\end{figure}

In this section, we briefly describe the joint iterative channel detection and LDPC code decoding algorithm proposed in \cite{JYAO2}. The system diagram of the iterative detection and decoding process is illustrated in Fig. \ref{figure2}. The channel detector and LDPC decoder exchange extrinsic binary information and work iteratively to decode the input binary sequence of the system.
\subsection{Iterative Channel Detector}
\begin{figure}[!t]
\centering
\includegraphics[width=4in]{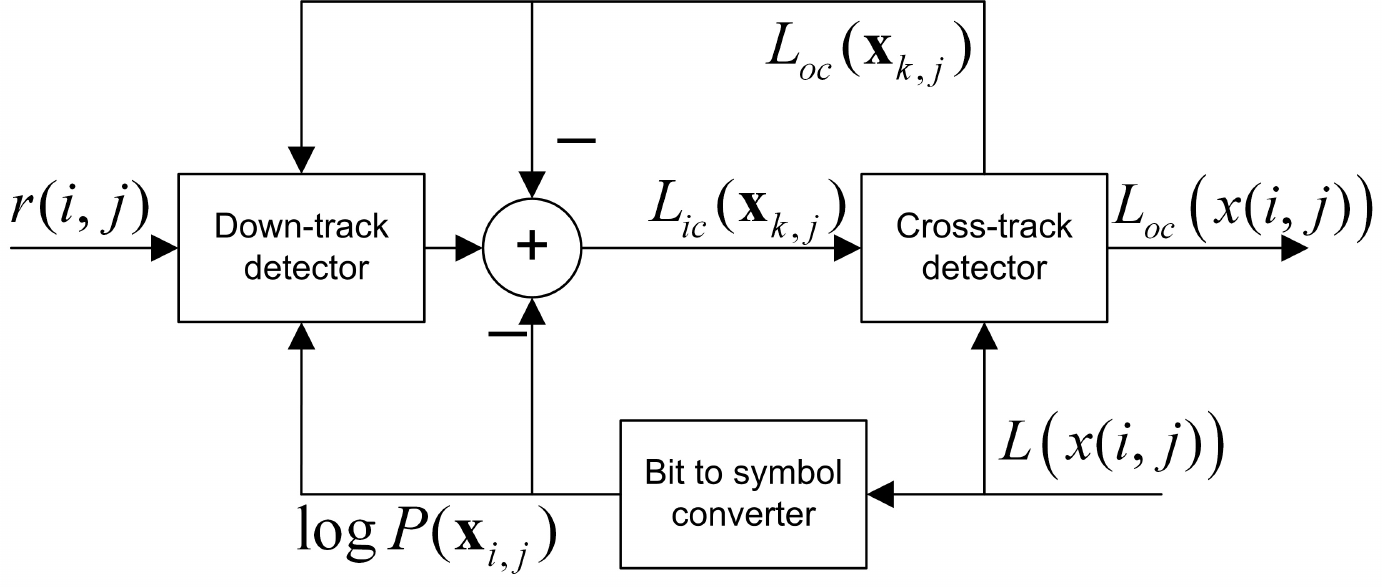}
% where an .eps filename suffix will be assumed under latex,
% and a .pdf suffix will be assumed for pdflatex; or what has been declared
% via \DeclareGraphicsExtensions.
\caption{Iterative detection scheme in the channel detector.}
\label{figure3}
\end{figure}

The channel detector mainly consists of two constituent detectors, namely, the down-track detector and cross-track detector as shown in Fig. \ref{figure3}. The down-track detector adopts the symbol-based BCJR algorithm with a trellis state defined as $S=[\textbf{x}_{i,k-N_{h}+1},\textbf{x}_{i,k-N_{h}+2},...,\textbf{x}_{i,k-1}]$. The log-domain symbol-based BCJR detection can be calculated as
\begin{eqnarray}
\label{6}
\gamma _{k}(S',S)=L_{id}(\textbf{x}_{i,k})-\frac{\left | r(i,k)-\sum\limits_{n=0}^{N_{h}-1}\textbf{x}_{i,k-N_{h}+1+n}\cdot \textbf{h}_{n} \right |^{2}}{2\sigma ^{2}}
\end{eqnarray}
\begin{eqnarray}
\label{7}
\alpha_{k}(S)={\max_{S'}}^*\left \{ \alpha _{k-1}(S')+\gamma _{k}(S',S) \right \}
\end{eqnarray}
\begin{eqnarray}
\label{8}
\beta_{k-1}(S')={\max_{S}}^*\left \{ \beta _{k}(S)+\gamma _{k}(S',S) \right \}
\end{eqnarray}
\begin{eqnarray}
\label{9}
L_{od}(\textbf{x}_{i,k})&=&\log \left ( P(\textbf{x}_{i,k}=U_{t}|r) \right ) \nonumber \\
        &=&{\max_{U_{t}}}^*\left \{  \alpha _{k-1}(S')+\gamma _{k}(S',S)+\beta _{k}(S)\right \}
\end{eqnarray}
where $L_{id}(\textbf{x}_{i,k})$ is the logarithmic $a~priori$ probability of the symbol $\textbf{x}_{i,k}$, the subscript implies $L_{id}(\textbf{x}_{i,k})$ is the input to the down-track detector. $U_{t}$ represents the $t$-th element from the symbol alphabet $U$ of size $2^{M_{H}}$, $\alpha _{k}(S)$ and $\beta _{k}(S)$ are the forward and backward state metrics, respectively. They are initialized as
\begin{eqnarray}
\label{initialization1}
\alpha_0(S)=\beta_{N_c+1}(S)=\left\{\begin{matrix}
&0&,&&\textup{state with all -1s}\\
&-\infty&,&&\textup{other states}
\end{matrix}\right..
\end{eqnarray}
$L_{od}(\textbf{x}_{i,k})$ is the generated $a~posteriori$ logarithm probability of the symbol $\textbf{x}_{i,k}$, the subscript implies it is the output of the down-track detector. The extrinsic information transferred from the down-track detector to the cross-track detector can thus be obtained as
\begin{eqnarray}
\label{inserteq1}
L_{ic}(\textbf{x}_{i,k})=L_{od}(\textbf{x}_{i,k})-L_{id}(\textbf{x}_{i,k})
\end{eqnarray}
where the subscript of $L_{ic}(\textbf{x}_{i,k})$ implies that it is the input information to the cross-track detector. The function ${\max}^{*}$ is defined as \cite{ref26}
\begin{eqnarray}
\label{10}
{\max}^{*}(x,y)&\triangleq&\log(e^{x}+e^{y})\nonumber\\
&=&\max(x,y)+\log(1+e^{-|x-y|}).
\end{eqnarray}
Due to the symmetries of the channel response matrix, the symbol $\textbf{x}_{i,k}$ is not fully distinguishable. Thus, the cross-track detector is needed to further detect the binary bits in the cross-track direction. The cross-track detector also adopts the trellis-based algorithm. A state in the trellis is defined as $S_{k}^{c}=[x(k-M_{h}+1,j),x(k-M_{h}+2,j),...,x(k-1,j)]$. Similar to the down-track detector, the log-domain cross-track detection algorithm can be implemented as
\begin{eqnarray}
\label{11}
\mathfrak{c}_k(S_{k}^c,S_{k+1}^c)=L_{ic}(\textbf{x}_{k,j})+\log (P(x(k,j)))
\end{eqnarray}
\begin{eqnarray}
\label{12}
\mathfrak{a}_{k}(S_{k+1}^c)={\max_{S_{k}^c}}^*\{\mathfrak{a}_{k-1}(S_{k}^c)+\mathfrak{c}_k(S_{k}^c,S_{k+1}^c)\}
\end{eqnarray}
\begin{eqnarray}
\label{13}
\mathfrak{b}_{k-1}(S_{k}^c)={\max_{S_{k+1}^c}}^*\{\mathfrak{b}_{k}(S_{k+1}^c)+\mathfrak{c}_k(S_{k}^c,S_{k+1}^c)\}
\end{eqnarray}
\begin{eqnarray}
\label{14}
L_{oc}(\textbf{x}_{k,j})=\mathfrak{a}_{k-1}(S_{k}^c)+\log(P(x(k,j)))+\mathfrak{b}_k(S_{k+1}^c).
\end{eqnarray}
\begin{eqnarray}
\label{15}
L_{oc}(x(k,j))&=&\log\left (  \frac{P(x(k,j)=1)}{P(x(k,j)=-1)}\right )\nonumber \\
&=&{\max_{V_1}}^*\{\mathfrak{a}_{k-1}(S_{k}^c)+\mathfrak{c}_k(S_{k}^c,S_{k+1}^c)+\mathfrak{b}_k(S_{k+1}^c) \}\nonumber \\
&~&-{\max_{V_{-1}}}^*\{\mathfrak{a}_{k-1}(S_{k}^c)+\mathfrak{c}_k(S_{k}^c,S_{k+1}^c)+\mathfrak{b}_k(S_{k+1}^c) \}
\end{eqnarray}
where $\log (P(x(k,j)))$ is the $a~priori$ information of the binary bit $x(k,j)$. The forward metric $ \mathfrak{a}_{k}(S_{k+1}^c)$ and backward metric $\mathfrak{b}_{k}(S_{k+1}^c)$ are initialized as
\begin{eqnarray}
\label{initialization2}
\mathfrak{a}_{0}(S_{1}^c)=\mathfrak{b}_{N_r+1}(S_{N_r+2}^c)=\left\{\begin{matrix}
&0&,&&\textup{state with all -1s}\\
&-\infty&,&&\textup{other states}
\end{matrix}\right..
\end{eqnarray}
$V_1$ represents the set of pairs $(S_k^c, S_{k+1}^c)$ that correspond to the event $x(i,j)=1$, and $V_{-1}$ is similarly defined. $L_{oc}(\textbf{x}_{k,j})$ and $L_{oc}(x(k,j))$ are the newly generated symbol-based and bit-based information, respectively. The subscripts imply that they are the output of the cross-track detector.

In order to make use of the binary information in the down-track detector, a bit-to-symbol converter should be implemented via
\begin{eqnarray}
\label{16}
\log P(\textbf{x}_{i,j})= \sum_{k=0}^{M_{h}-1}\log P(x(i-k,j)).
\end{eqnarray}
The above symbol-based soft information $\log P(\textbf{x}_{i,j})$ and $L_{oc}(\textbf{x}_{k,j})$ in (\ref{16}) will be used to update the $a~priori$ information of $L_{id}(\textbf{x}_{k,j})$ in (\ref{6}) as shown in Fig. \ref{figure3}.

\subsection{LDPC Decoder}
In the $t$-th round iteration between the channel detector and LDPC decoder, let the deinterleaved extrinsic logarithmic likelihood ratios (LLRs) from the channel detector be denoted by $\{ L^t_{ext1}(n),n=1,...,N \}$. The LDPC decoder adopts the log-domain sum-product algorithm which mainly consists of the variable-to-check message update and the check-to-variable message update \cite{ref27}. In the $l$-th iteration inside the LDPC decoder, the LLR $z_{mn}^{l,t}$ of the $n$-th bit which is sent from the variable node $n$ to check node $m$ can be calculated as
\begin{eqnarray}
\label{17}
z_{mn}^{l,t}=L^t_{ext1}(n)+\sum_{j\in D_v(n)\backslash m}q_{jn}^{l-1,t}
\end{eqnarray}
where $ D_v(n)\backslash m$ represents all the check nodes connected to the variable node $n$ except for the check node $m$ and  $q_{mn}^{l,t}$ is the LLR of bit $n$ which is sent from the check node $m$ to variable node $n$. The update of $q_{mn}^{l,t}$ can be calculated as
\begin{eqnarray}
\label{18}
\tanh[q_{mn}^{l,t}/2]=(-1)^{d(m)}\cdot \prod_{k\in D_c(m)\backslash n}\tanh[z_{mk}^{l,t}/2]
\end{eqnarray}
where $D_c(m)\backslash n$ represents all the variable nodes connected to the check node $m$ except for the variable node $n$. The output LLR of the $n$-th bit of the LDPC decoder after $t$ rounds of iterations between the channel detector and LDPC decoder can be calculated via
\begin{eqnarray}
\label{19}
L^t_{c}(n)=L^t_{ext1}(n)+\sum_{j\in D_v(n)}q_{jn}^{l,t}
\end{eqnarray}
where $D_v(n)$ represents all the check nodes connected to the variable node $n$. The extrinsic information from the LDPC decoder to the channel detector in the $t$-th iteration can be calculated via
\begin{eqnarray}
\label{20}
L^t_{ext2}(n)=\sum_{j\in D_v(n)}q_{jn}^{l,t}.
\end{eqnarray}

\subsection{The Full Message-Passing Algorithm}
\begin{figure}[!t]
\centering
\includegraphics[width=5in]{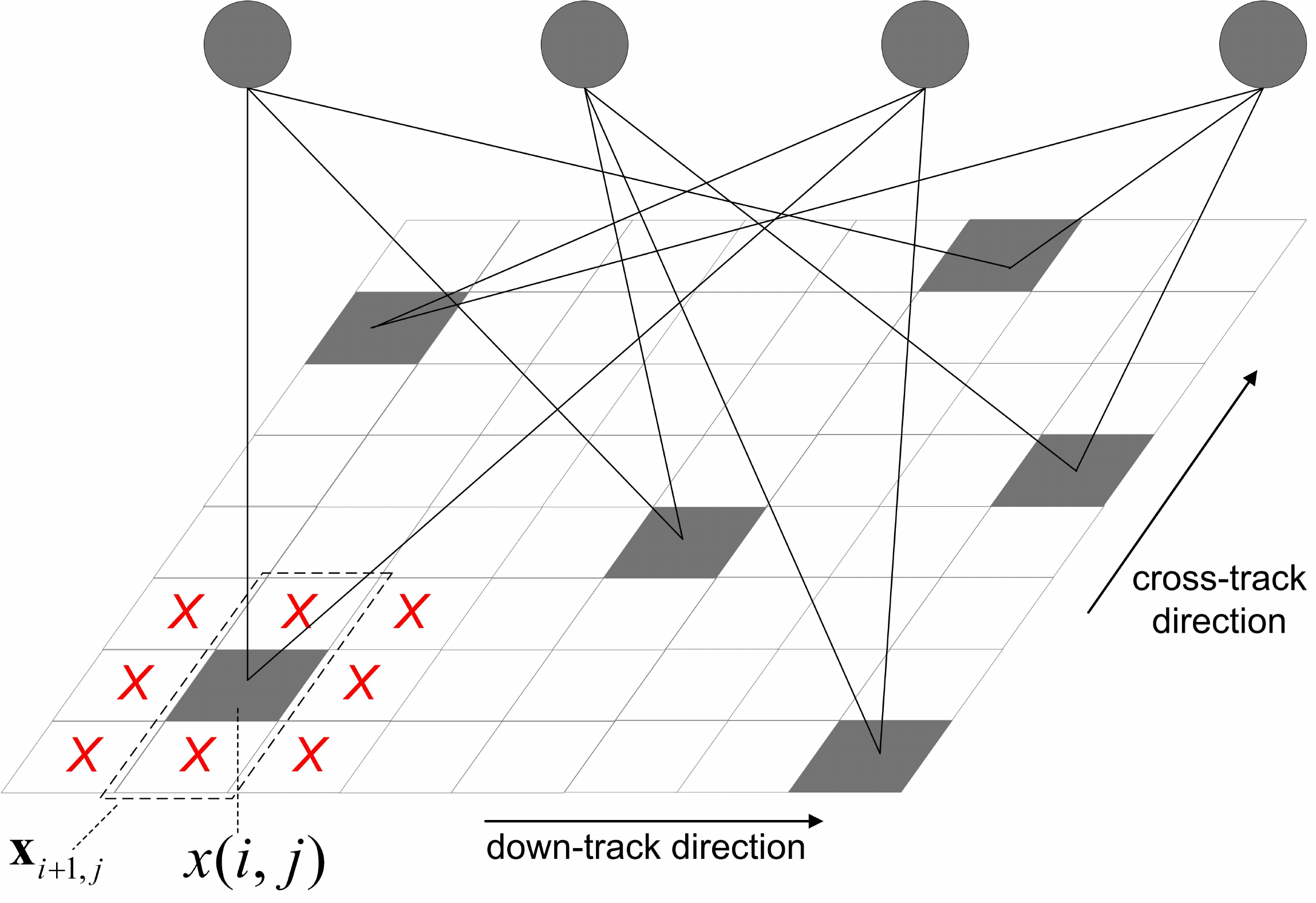}
% where an .eps filename suffix will be assumed under latex,
% and a .pdf suffix will be assumed for pdflatex; or what has been declared
% via \DeclareGraphicsExtensions.
\caption{Factor graph representation of the joint detection and decoding algorithm.}
\label{figure4}
\end{figure}

Figure \ref{figure4} shows a factor graph representation of the system. The dark grids represent the variable nodes and the dark circles represent the check nodes. The variable nodes connected to the same check node are constrained by a parity-check equation. The grids with crosses around $x(i,j)$ represent all the binary signals that can interfere with $x(i,j)$ and they form an interference region of $x(i,j)$. Similarly, $x(i,j)$ can also interfere with all the bits within its interference region. In this figure, $M_h=N_h=3$ and each variable node is checked by 2 check nodes and each check node is connected to 3 variable nodes. The joint channel detection and LDPC codes decoding algorithm is executed iteratively. Here we assume that the simplest stopping criterion is adopted, i.e., the iterations are conducted for a pre-set number of times. The full message-passing algorithm in the joint channel detection and LDPC codes decoding can be summarized as

\begin{itemize}
	\item Initialization

\begin{enumerate}
	\item receive all the channel output $ \{ r(j,k),j=1,...,N_r,k=1,...,N_c \}$;
	\item set all the $a~priori$ and extrinsic information $\log P(\textbf{x}_{j,k})$, $ \log (P(x(j,k)))$, $ q_{mn}^{0,t}$ to zeros;
	\item set $t=1$.
\end{enumerate}

	\item Joint iterative detection/decoding
	
	\begin{enumerate}
	\item execute the channel detection algorithm (\ref{6})-(\ref{15}) iteratively for $I_{det}$ times;
	\item deinterleave the soft information from the channel detector and obtain the extrinsic information from the channel detector to the LDPC decoder $L^t_{ext1}(n)$;
	\item compute the variable-to-check messages $z_{mn}^{l,t}$ and check-to-variable messages $q_{mn}^{l,t}$ iteratively for $I_{c}$ times;
	\item compute the extrinsic information from the LDPC decoder to the channel detector $L^t_{ext2}(n)$. Interleave the information $L^t_{ext2}(n)$ and use it to update the $a~priori$ information in the channel detector;
	\item if $t<I_{out}$, repeat the processes 1)- 4). Increment $t$ by 1.
  \end{enumerate}

	\item Decode
	\begin{enumerate}
	\item compute the estimated bits using $\widehat{s}(n)=\left (\textup{sign} (L^t_{c}(n))+1  \right )/2$.
  \end{enumerate}
\end{itemize}

\section{Concentration, density evolution and threshold calculation}
In this section, we will first explain the message-flow neighborhood structures of the iterative detection system. We will then prove a concentration statement for every possible input sequence, and we will show that the average performance of the system is closely concentrated around the system performance when the input sequence is independent and uniformly distributed (i.u.d.). Based on the concentration result, we will introduce the density evolution method to track the message densities in the iterations. Finally, using the density evolution technique, we can calculate a threshold which can be considered as the performance limit of the system.
\subsection{Message-Flow Neighborhoods}
$Assumption~1$: In the channel detection algorithm, the soft decision of a certain bit is mainly dependent on the information of its finite-size neighboring bits, i.e., the value of $L(x(i,j))$  is calculated based on the information corresponding to the bits $\{ x(k,l): k=i-F_c,...,i+F_c, l=j-F_d,...,j+F_d \}$ with $ F_c<+\infty$ and $ F_d<+\infty $.

In order to verify this assumption, we will introduce the windowed-version channel detector. In the original channel detector, all the received signals are processed and the full page of data is detected as a whole. In contrast to the original channel detector, for the detection of the bit $x(i,j)$, the windowed-version channel detector does not require the whole page of received signals but only require the received signals that are adjacent to the bit $x(i,j)$, i.e., the received signals $\{ r(k,l): k=i-F_c-M_h+1,...,i+F_c+M_h-1, l=j-F_d-N_h+1,...,j+F_d+N_h-1\}$. Since this finite size region is not surrounded by $-1$s, the forward and backward metrics cannot be initialized to the zero state. Instead, they are initialized as $\alpha_{j-F_d}(S)= \beta_{j+F_d+N_h-1}(S)=-M_h(N_h-1)\cdot \log(2)$, for $S=0,1,...,2^{M_h(N_h-1)}-1$ and  $\mathfrak{a}_{i-F_c}(S^c)= \mathfrak{b}_{i+F_c+M_h-1}(S^c)=-(M_h-1) \cdot \log(2)$, for $S^c=0,1,...,2^{M_h-1}-1$. The detection algorithm of the windowed-version detector inside this region is the same as the original detector.

\begin{figure}[!t]
\centering
\includegraphics[width=5in]{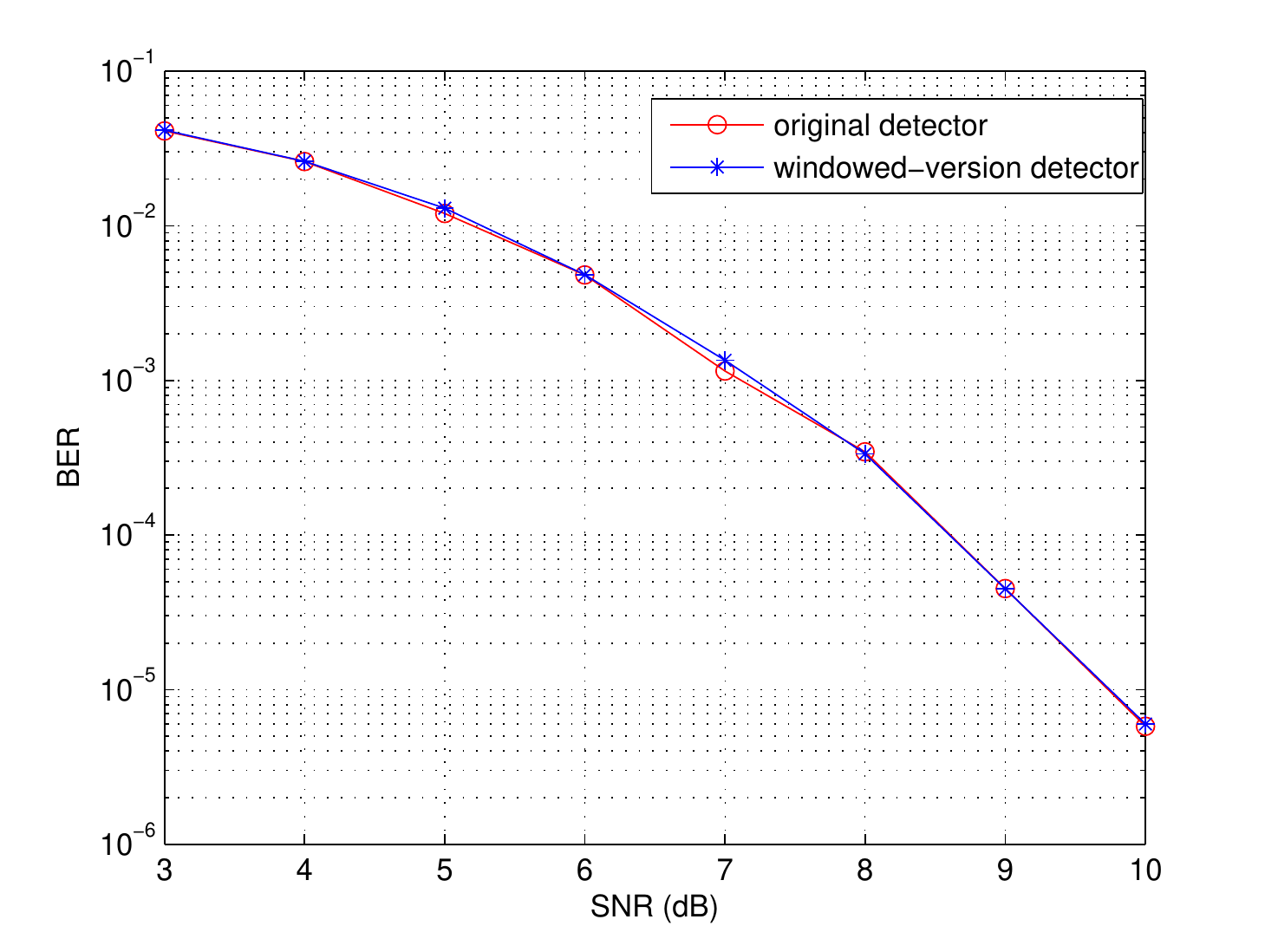}
% where an .eps filename suffix will be assumed under latex,
% and a .pdf suffix will be assumed for pdflatex; or what has been declared
% via \DeclareGraphicsExtensions.
\caption{Performance comparisons of the original channel detector and the windowed-version channel detector.}
\label{figure5}
\end{figure}

Figure \ref{figure5} shows the simulated bit-error rate (BER) curves versus SNR for both the original channel detector and the windowed-version channel detector. The channel response matrix used in the simulation is \cite{MCMC}
\begin{eqnarray}
\label{21}
\textbf{H}_A=\left(
\begin{array}{ccc}
 0.050684 & 0.21273 & 0.050684 \\
 0.23825 & 1 & 0.23825 \\
 0.050684 & 0.21273 & 0.050684
\end{array}
\right).
\end{eqnarray}

In the simulation, we set $F_c=F_d=5$ for the windowed-version channel detector. It can be observed that the windowed-version channel detector performs nearly identical to the original channel detector. Thus, it can be concluded that the soft decision of $x(i,j)$ based on the information of its neighboring bits cannot be improved when more information becomes available, which verifies Assumption 1. In the following discussions, we will assume Assumption 1 is held.

\begin{figure}[!t]
\centering
\includegraphics[width=3.5in]{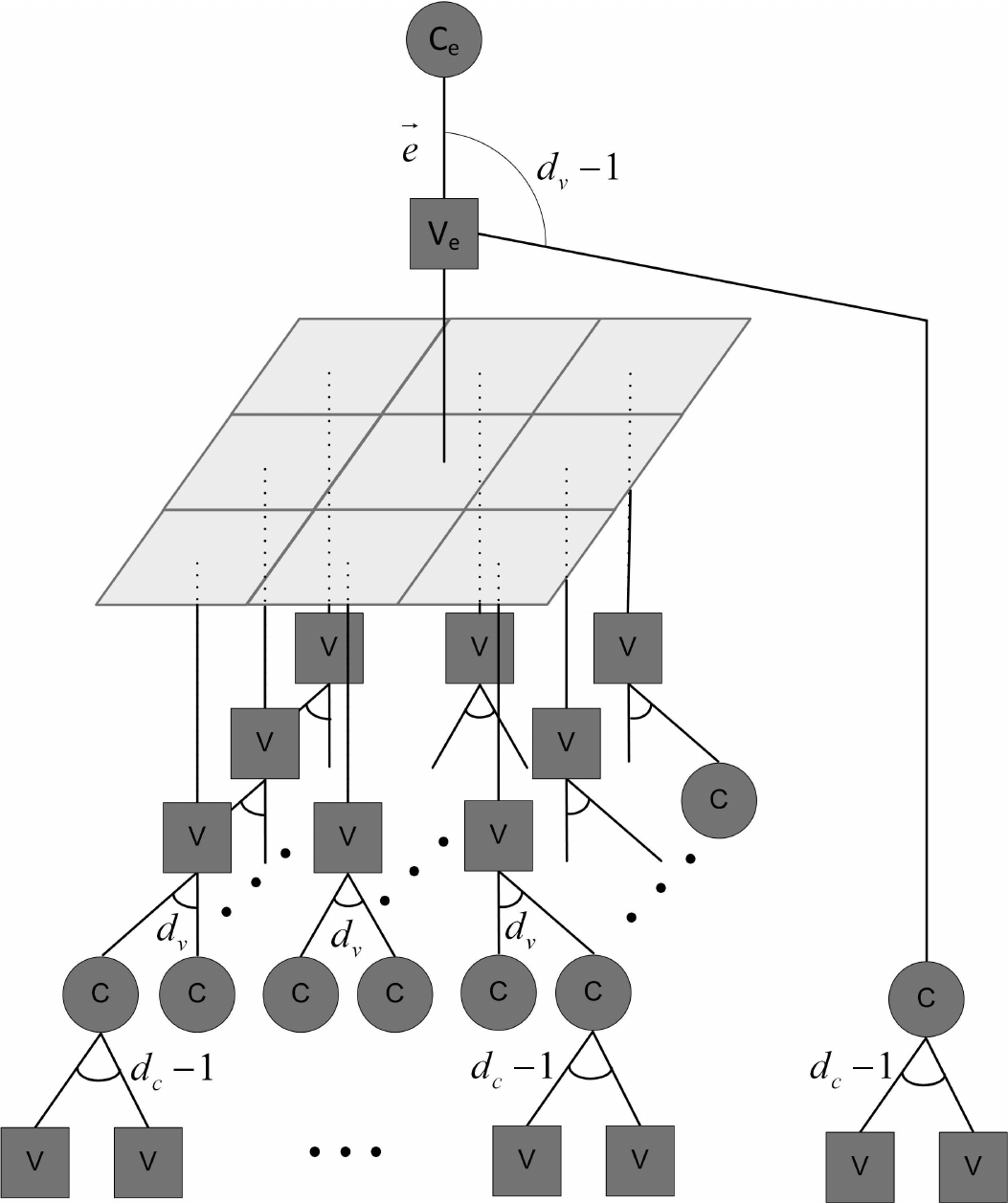}
% where an .eps filename suffix will be assumed under latex,
% and a .pdf suffix will be assumed for pdflatex; or what has been declared
% via \DeclareGraphicsExtensions.
\caption{Message flow neighborhood of a variable node to a check node with $t=1$, $I_c=1$ and $F_c=F_d=1$.}
\label{figure6}
\end{figure}

For clarity of presentation, we consider regular LDPC codes, where each variable node has degree $d_v=d_v^{max}$ and each check node has degree $d_c=d_c^{max}$. Consider an edge $\vec{e}$ that connects the variable node $V_e$ to the check node $C_e$ at the end of the $t$-th iteration between channel detector and LDPC decoder. The message-flow neighborhood of depth $t$ of the edge $\vec{e}$, denoted by $\mathbb{N}_{\vec{e}}^t$, is a subgraph that consists of the two nodes $V_e$ and $C_e$, the edge $\vec{e}$, and all nodes and edges that contribute to the computation of the message $z_{C_eV_e}^{I_{c},t}$ passed from $V_e$ to $C_e$. Figure \ref{figure6} shows a message-flow neighborhood with $t=1$, $I_{c}=1$ and $F_c=F_d=1$. For a depth-1 message-flow neighborhood, the message $z_{C_eV_e}^{I_{c},t}$ is directly dependent on its $ I_{c}-1$ stages of check nodes and variable nodes. The information of the last-stage variable nodes is obtained from the channel detector. From Assumption 1, in the channel detector, the detection of any bit is dependent on its $4F_cF_d-1$ neighboring bits. Each of these neighboring bits also corresponds to a variable node which is further constrained by $I_{c}$ stages of check nodes and variable nodes. The message-flow neighborhood of depth $t$ can be obtained similarly by branching out the last-stage variable-to-check edges of the depth-1 neighborhood.

$Theorem~1$: When the code graph is chosen uniformly at random from the ensemble of all graphs, then for sufficiently large $N$, the message-flow neighborhood $\mathbb{N}_{\vec{e}}^{t^*}$ will form a tree with the probability that
\begin{eqnarray}
\label{22}
\Pr \{\mathbb{N}_{\vec{e}}^{t^*}\textup{ is not tree-like}\}\leq \gamma /N
\end{eqnarray}
for some constant $\gamma$, where $t^*$ is a fixed depth which represents the number of iterations between channel detector and LDPC decoder.

$Proof$: In a message-flow neighborhood of depth $t$, there are, in total,
\begin{eqnarray}
\label{23}
Q_v^t&:=&\sum _{i=0}^{I_{c}-1}[(d_v-1)(d_c-1)]^i\nonumber\\
&+&[(d_v-1)(d_c-1)]^{I_c-1}\cdot \sum _{j=1}^t\{(4F_cF_d-1)[1+d_v\sum _{i=1}^{I_c-1}(d_v-1)^{i-1}(d_c-1)^i]\}^j
\end{eqnarray}
variable nodes and
\begin{eqnarray}
\label{24}
Q_c^t&:=&1+(d_v-1)\sum _{i=0}^{I_{c}-2}[(d_v-1)(d_c-1)]^i\nonumber\\
&+&[(d_v-1)(d_c-1)]^{I_c-1}\cdot \sum _{j=1}^t\{(4F_cF_d-1)d_v\sum _{i=1}^{I_c-1}[(d_v-1)(d_c-1)]^{i-1}\}^j
\end{eqnarray}
check nodes. Assume that $\mathbb{N}_{\vec{e}}^t$ is tree-like, with $t+1<t^*$. When we branch out the neighborhood $\mathbb{N}_{\vec{e}}^t$ to form the neighborhood $\mathbb{N}_{\vec{e}}^{t+1}$, it can be shown that the probability that the newly revealed check nodes do not create any loop can be lower bounded by $(1-Q_c^{t^*}/M)^{Q_c^{t+1}-Q_c^{t}}$, with $M=N-K$. Similarly, the probability that the newly revealed variable nodes do not create any loop can be lower bounded by $(1-Q_v^{t^*}/N)^{Q_v^{t+1}-Q_v^{t}}$. Thus, the probability that $\mathbb{N}_{\vec{e}}^{t^*}$ is tree-like is lower bounded by
\begin{eqnarray}
\label{insert1}
(1-Q_v^{t^*}/N)^{Q_v^{t^*}}(1-Q_c^{t^*}/M)^{Q_c^{t^*}}.
\end{eqnarray}
Note that $Md_c=Nd_v=N_e$. Therefore, for sufficiently large $N$, it has
\begin{eqnarray}
\label{25}
\Pr \{\mathbb{N}_{\vec{e}}^{t^*}\textup{ is not tree-like}\}\leq \frac{(Q_v^{t^*})^2+\frac{d_c}{d_v} (Q_c^{t^*})^2}{N}=\gamma /N
\end{eqnarray}
where $\gamma$ is a constant dependent on $t$, $I_c$, $d_v$, $d_c$, $F_cF_d$, but not on $N$.

\subsection{Concentration Theorems}
For sufficiently large $N$, the message-flow neighborhood will form a tree which can be specified by the binary bits of variable nodes in the tree. Denote $N_v(t)$ as the number of possible fillings of binary bits in the tree-like neighborhood of depth $t$. Each possible filling is referred to as a message-flow neighborhood type. We index these neighborhoods as
\begin{eqnarray}
\label{insert2}
{\theta} _i\in \{0,1\}^{N_v(t)},\textup{ where }1\leq i\leq N_v(t).
\end{eqnarray}

Denote $c_{{\theta}}$ as the bipolar bit corresponding to the variable node $V_e$ at the top of the message-flow neighborhood of type ${\theta}$. Define $p^t({{\theta}})$ as the probability that a tree of type ${\theta}$ and depth $t$ delivers an incorrect $c_{{\theta}}$, i.e.,
\begin{eqnarray}
\label{26}
p^t({{\theta}})=\Pr (z_{C_eV_e}^{I_{c},t}\cdot c_{{\theta}}<0|\textup{tree type }{\theta}).
\end{eqnarray}

Denote $\textbf{c}$ as a particular sequence of transmitted bits. For a given edge $\vec{e}$ whose message-flow neighborhood of depth $t$ is tree-like, let $p^t(\textbf{c})$ be the expected number of incorrect messages passed along this edge at the $t$-th iteration when $\textbf{c}$ is transmitted. Thus, $p^t(\textbf{c})$ can be expressed as
\begin{eqnarray}
\label{27}
p^t(\textbf{c})=\sum _{i=1}^{N_v(t)}p^t({{\theta} _i})\Pr({\theta}_i|\textbf{c}) .
\end{eqnarray}

$Theorem~2$: Denote $e^t(\textbf{c})$ as the number of erroneous variable-to-check messages when $\textbf{c}$ is the transmitted codeword and $t$ rounds of the message-passing algorithm have been implemented. Assume the code graph is chosen uniformly at random from the ensemble of all graphs. Let $N_e$ be the total number of variable-to-check edges in the graph. For any $\epsilon >0$, there exists a positive constant $\beta$, such that if $N$ is sufficiently large, then
\begin{eqnarray}
\label{28}
\Pr\left(\left| \frac{e^t(\textbf{c})}{N_e} -p^t(\textbf{c})\right |>\epsilon \right)\leq 2e^{-\beta \epsilon ^2N}.
\end{eqnarray}

Denote $\textbf{c}_{\textup{i.u.d.}}$ as an i.u.d. random sequence. For a given edge $\vec{e}$ whose message-flow neighborhood of depth $t$ is tree-like, let $p^t_{\textup{i.u.d.}}$ be the expected number of incorrect messages passed along this edge at the $t$-th iteration when an i.u.d. random sequence is transmitted. The value $p^t_{\textup{i.u.d.}}$ can be expressed as
\begin{eqnarray}
\label{34}
p^t_{\textup{i.u.d.}}&=&\textbf{E}\left [ p^t(\textbf{c}_{\textup{i.u.d.}}) \right ]=\sum _{j=1}^{2^{N}}2^{-N}p^t(\textbf{c}_j)\nonumber\\
&=&\sum _{j=1}^{2^N}2^{-N}\sum _{i=1}^{N_v(t)}p^t({{\theta} _i})\Pr({\theta}_i|\textbf{c}_j)\nonumber\\
&=&\sum _{i=1}^{N_v(t)}p^t({\theta}_i)\sum _{j=1}^{2^N}2^{-N}\Pr({\theta}_i|\textbf{c}_j)\nonumber\\
&=&\sum _{i=1}^{N_v(t)}p^t({\theta}_i)\Pr({\theta}_i|\textbf{c}_{\textup{i.u.d.}})\nonumber\\
&=&\sum _{i=1}^{N_v(t)}p^t({\theta}_i)\frac{1}{N_v(t)}.
\end{eqnarray}
The last step of the above equation comes from the fact that all neighborhood types are equally probable when the transmitted sequence is i.u.d..

$Theorem~3$: Denote $e^t(\textbf{c}_{\textup{i.u.d.}})$ as the number of erroneous variable-to-check messages when an i.u.d. codeword $\textbf{c}_{\textup{i.u.d.}}$ is transmitted and $t$ rounds of the message-passing algorithm have been implemented. Assume the code graph is chosen uniformly at random from the ensemble of all graphs. For any $\epsilon >0$, there exists a positive constant $\beta$ such that if $N$ is sufficiently large, then
\begin{eqnarray}
\label{35}
\Pr\left(\left| \frac{e^t(\textbf{c}_{\textup{i.u.d.}})}{N_e} -p^t_{\textup{i.u.d.}}\right |>\epsilon \right)\leq 2e^{-\beta \epsilon ^2N}.
\end{eqnarray}

Theorems 2 and 3 can be proved similarly by following the proofs of Theorems 1 and 2 in \cite{ref14}. Theorem 3 states that if an i.u.d. random sequence is transmitted, then the probability of a variable-to-check message being incorrect after $t$ rounds of the message-passing decoding algorithm is highly concentrated around the probability $p^t_{\textup{i.u.d.}}$ in which the depth-$t$ message-flow neighborhood is tree-like. Thus, we can use $p^t_{\textup{i.u.d.}}$ to evaluate the performance of the system with i.u.d. transmitted sequence.

\subsection{Density Evolution and Threshold Calculation}
Based on the tree-like message-flow neighborhood, we can analyze the decoding algorithm by tracking the evolution of the message densities during the iterations. This method was used in \cite{ref8} to analyze the performance of LDPC codes over memoryless channels, where it was called the density evolution. Let $\tau^t(n)$ represent the correct extrinsic LLR of the $n$-th binary bit from the channel detector to the LDPC decoder after $t$ rounds of the message-passing algorithm, i.e., $\tau^t(n)=L^t_{ext1}(n)\cdot c(n)$. Let $f$ represent the probability density function (pdf) of the corresponding variable. The evolution of the message densities of $\tau^t(n)$ through channel detector can be expressed as
\begin{eqnarray}
\label{38}
f^t_{\tau}= \xi _c(f^{t-1}_{L_{ext2} },f_N)
\end{eqnarray}
where $\xi_c(.)$ is symbolic notation of the message density evolution in the channel detector. Since no closed-form expression of $\xi_c(.)$ can be obtained, it is calculated using Monte Carlo techniques. The evolution of average correct message densities inside the LDPC decoder can be expressed as
\begin{eqnarray}
\label{39}
f^{l,t}_{z}= f^t_{\tau}*\lambda (f^{l-1,t}_{q})
\end{eqnarray}
\begin{eqnarray}
\label{40}
f^{l,t}_{q}=\rho \left [\xi \left ( f^{l,t}_{z} \right )\right ]
\end{eqnarray}
\begin{eqnarray}
\label{41}
f^t_{L_{ext2}} = f^t_{\tau}*\bar{\lambda }\left ( f^{l,t}_{q} \right )
\end{eqnarray}
where $\xi(.)$ is a symbolic notation of the average message density obtained by evolving the density $f_{z^{l+1}}$ through a check node, and $\bar{\lambda }(x)=\sum _{i=1}^{d_v^{max}}\lambda_ix^i/[i\int _0^1\lambda(x)dx]$. From (\ref{34}), the value $p^t_{\textup{i.u.d.}}$ can be calculated as
\begin{eqnarray}
\label{42}
p^t_{\textup{i.u.d.}}&=& \sum _{i=1}^{N_v(t)}p^t({\theta}_i)\frac{1}{N_v(t)}\nonumber\\
&=&\sum _{i=1}^{N_v(t)}\frac{1}{N_v(t)}\int _{-\infty}^0f^{I_c,t}_{z|{\theta }_i}(\tau|{{\theta }_i})c_{{\theta }_i}d\tau\nonumber\\
&=&\int _{-\infty}^0\left[ \sum _{i=1}^{N_v(t)}\frac{1}{N_v(t)}f^{I_c,t}_{z|{\theta }_i}(\tau|{{\theta }_i})c_{{\theta }_i}\right ]d\tau\nonumber\\
&=&\int _{-\infty}^0f^{I_c,t}_z(\tau)d\tau.
\end{eqnarray}

Since the performance of the system degrades monotonously as the standard deviation $\sigma$ of the AWGN increases for a given channel response matrix, the threshold corresponds to the supremum of all the values of $\sigma$ such that the fraction of incorrect messages converges to zero as the code length and number of iterations tend to infinity. For the turbo-equalized system in which $t$ rounds of outer iterations are implemented between the channel detector and LDPC decoder and $I_c$ rounds of iterations are implemented inside the LDPC decoder per outer iteration, the threshold is defined as
\begin{eqnarray}
\label{43}
\sigma _{te}=\sup\left \{ \sigma :\lim _{t,I_c,N\rightarrow \infty}p^t_{\textup{i.u.d.}} \rightarrow 0\right \}.
\end{eqnarray}

For the non-turbo equalized system in which there is no outer iteration between the channel detector and LDPC decoder but only $l$ iterations inside the LDPC decoder,  the threshold is defined as
\begin{eqnarray}
\label{44}
\sigma _{non-te}=\sup\left \{ \sigma :\lim _{l,N\rightarrow \infty}\int _{-\infty}^0f^{l,1}_z(\tau)d\tau \rightarrow 0\right \}.
\end{eqnarray}

\section{Numerical results and discussions}
In this section, we will present the simulation results and the thresholds of LDPC coded 2D interference system detected by JIDDS. For comparisons, we will also include the simulation results and thresholds of LDPC codes over an AWGN channel \cite{ref8} and LDPC coded 2D interference system detected by a suboptimal strip-wise algorithm (SSWA) \cite{ref23}. The SSWA is called MAP algorithm in [28] because it is an optimal detection algorithm when only signals of three adjacent tracks are processed to detect the bit sequence in the middle track. For the SSWA, LDPC code is applied to the middle track and iterations are introduced between the detector and the LDPC decoder. Thus, the threshold for the SSWA can be comprehended as the threshold of LDPC coded 2D interference system where three tracks are used to detect the middle track. Note that the thresholds of SSWA are obtained by blindly adopting the same method as introduced in the previous section, but the concentration results are not proved.

For the calculation of thresholds for turbo-equalized system, we adopt a dynamic method. We first fix an error rate step $P_{ers}$ (e.g., $P_{ers}=10^{-6}$). In the $t$-th outer iteration, the iterations inside the LDPC decoder will proceed until the performance improvement achieved by the next iteration is smaller than the error rate step, i.e.,
\begin{eqnarray}
\label{45}
\int _{-\infty}^0f^{l+1,t}_z(\tau)d\tau-\int _{-\infty}^0f^{l,t}_z(\tau)d\tau<P_{ers}.
\end{eqnarray}
If (\ref{45}) is satisfied, then the iterations inside the LDPC decoder cease and the system moves to the $(t+1)$-th outer iteration. This dynamic method can save some computations while generating almost identical results in the calculation of threshold. Since it has been shown in \cite{JYAO2} that there is hardly any performance improvement by introducing more iterations when $I_{det}>3$, we fix $I_{det}=3$.

Table \ref{Table1} shows the evolution of $p^t_{\textup{i.u.d.}}$ during the iterations for various parameters. Besides the channel response matrix $\textbf{H}_A$, here we consider another channel response matrix \cite{MCMC}
\begin{eqnarray}
\label{H_B}
\textbf{H}_B=\left(
\begin{array}{ccc}
 0.0035638 & 0.14843 & 0.0035638 \\
 0.013382 & 0.55733 & 0.013382 \\
0.0035638 & 0.14843 & 0.0035638
\end{array}
\right).
\end{eqnarray}
It can be observed that for the given parameters, the error probability $p^t_{\textup{i.u.d.}}$ becomes smaller and smaller during the iterations and zero-error probability can be achieved within 5 iterations. However, if the standard deviation $\sigma$ exceeds the threshold, the error probability will be stuck at a certain value and will not evolve as the iteration process proceeds. Table \ref{Table2} shows the thresholds of LDPC coded 2D interference system detected by JIDDS and SSWA algorithm as well as thresholds of LDPC codes over an AWGN channel. We denote the threshold of LDPC code over AWGN channel and SSWA as $\sigma_{AWGN}$ and $\sigma_{SSWA}$, respectively in the table. For the 2D interference channel, the threshold is normalized by a factor $|\textbf{H}|=|\sum_{i=0}^{M_h-1}\sum_{j=0}^{N_h-1}\left ( h(i,j) \right )^2|^{1/2}$. If the standard deviation $\sigma$ is below the calculated threshold values, then any arbitrarily small target error probability can be achieved by using a sufficiently long code with sufficient number of iterations. It can be observed that as the code rate increases, the threshold decreases. Since the interference level of channel $\textbf{H}_B$ is smaller than that of $\textbf{H}_A$, the thresholds of channel $\textbf{H}_B$ is larger than that of $\textbf{H}_A$ when the system is detected by JIDDS. The JIDDS always has a larger threshold than the SSWA and the performance gap becomes larger as the code rate increases. This is because the SSWA only processes three tracks to detect the data sequence in the middle track, while the JIDDS detects the data in two directions and processes the information from more than three tracks.

\begin{figure}[!t]
\centering
\includegraphics[width=5in]{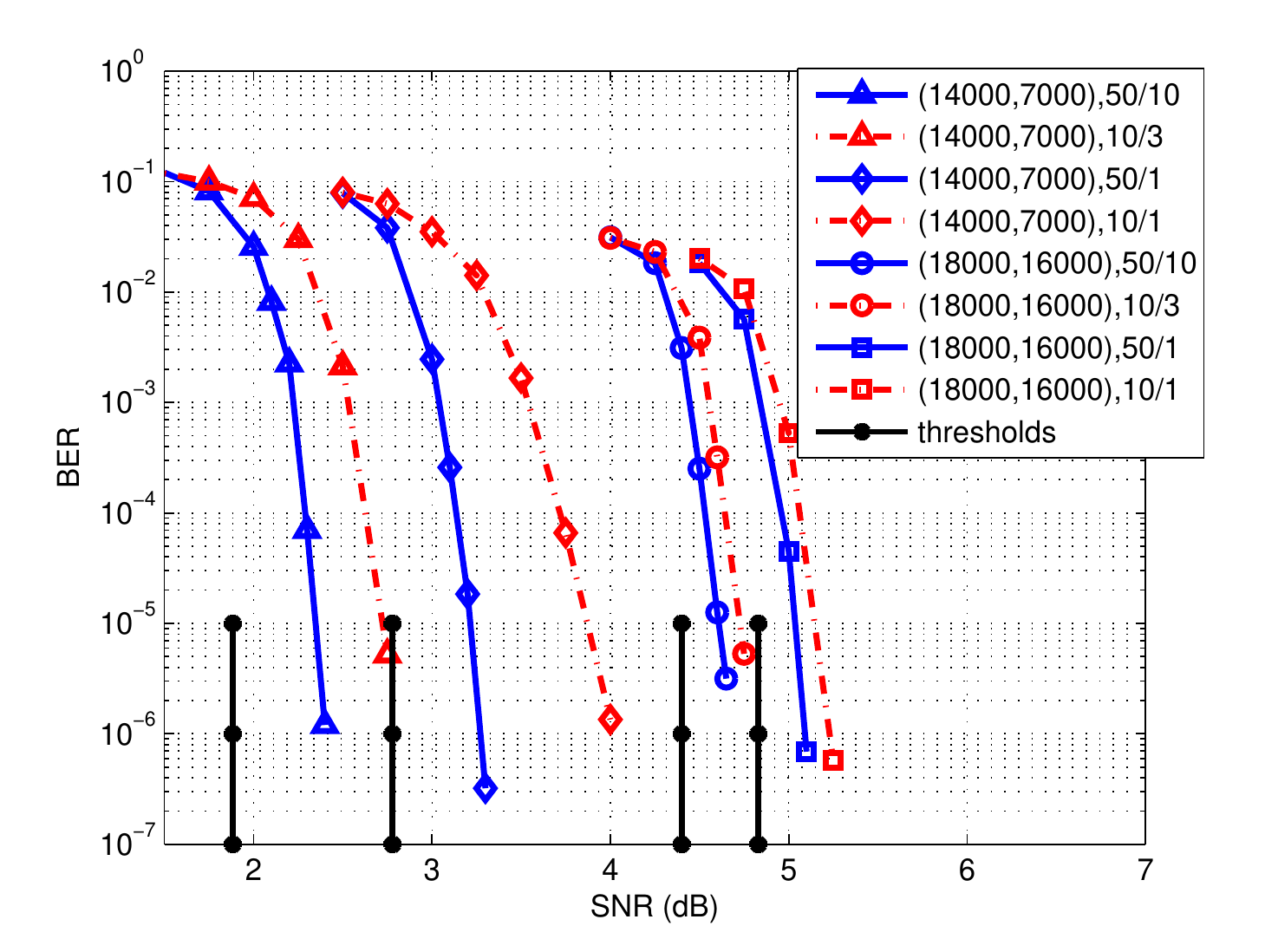}
% where an .eps filename suffix will be assumed under latex,
% and a .pdf suffix will be assumed for pdflatex; or what has been declared
% via \DeclareGraphicsExtensions.
\caption{Simulation results and thresholds for turbo-equalized and non-turbo equalized JIDDS with different LDPC codes and different number of iterations.}
\label{figure7}
\end{figure}

\begin{figure}[!t]
\centering
\includegraphics[width=5in]{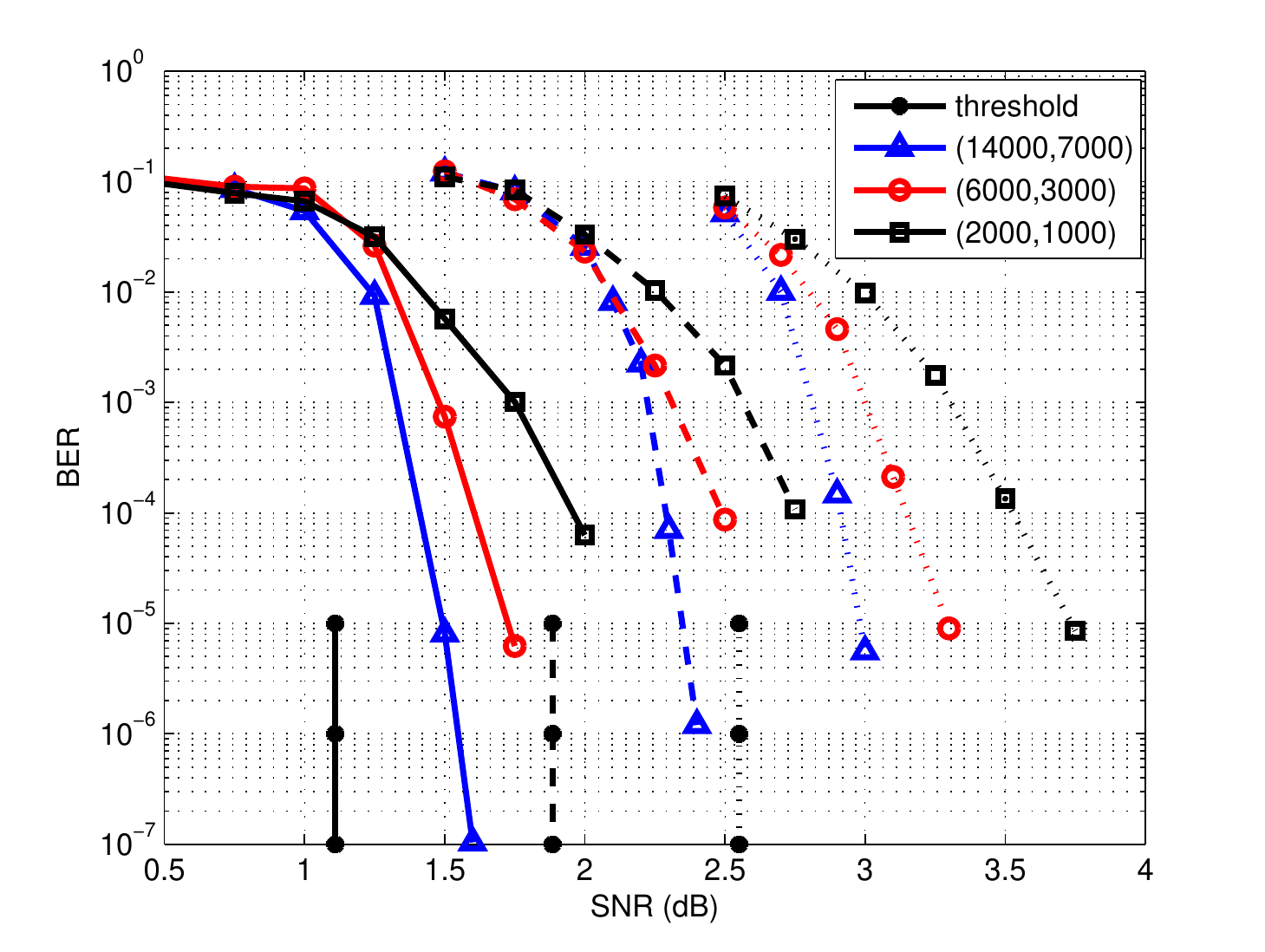}
% where an .eps filename suffix will be assumed under latex,
% and a .pdf suffix will be assumed for pdflatex; or what has been declared
% via \DeclareGraphicsExtensions.
\caption{Simulation results and thresholds for the LDPC codes with $d_v=3$, $d_c=6$ and $R=0.5$. The solid curves correspond to the LDPC codes over AWGN channel. The dashed curves correspond to the LDPC coded 2D interference system detected by JIDDS with turbo equalization. The dotted curves correspond to the LDPC coded 2D interference system detected by SSWA.}
\label{figure8}
\end{figure}

\begin{figure}[!t]
\centering
\includegraphics[width=5in]{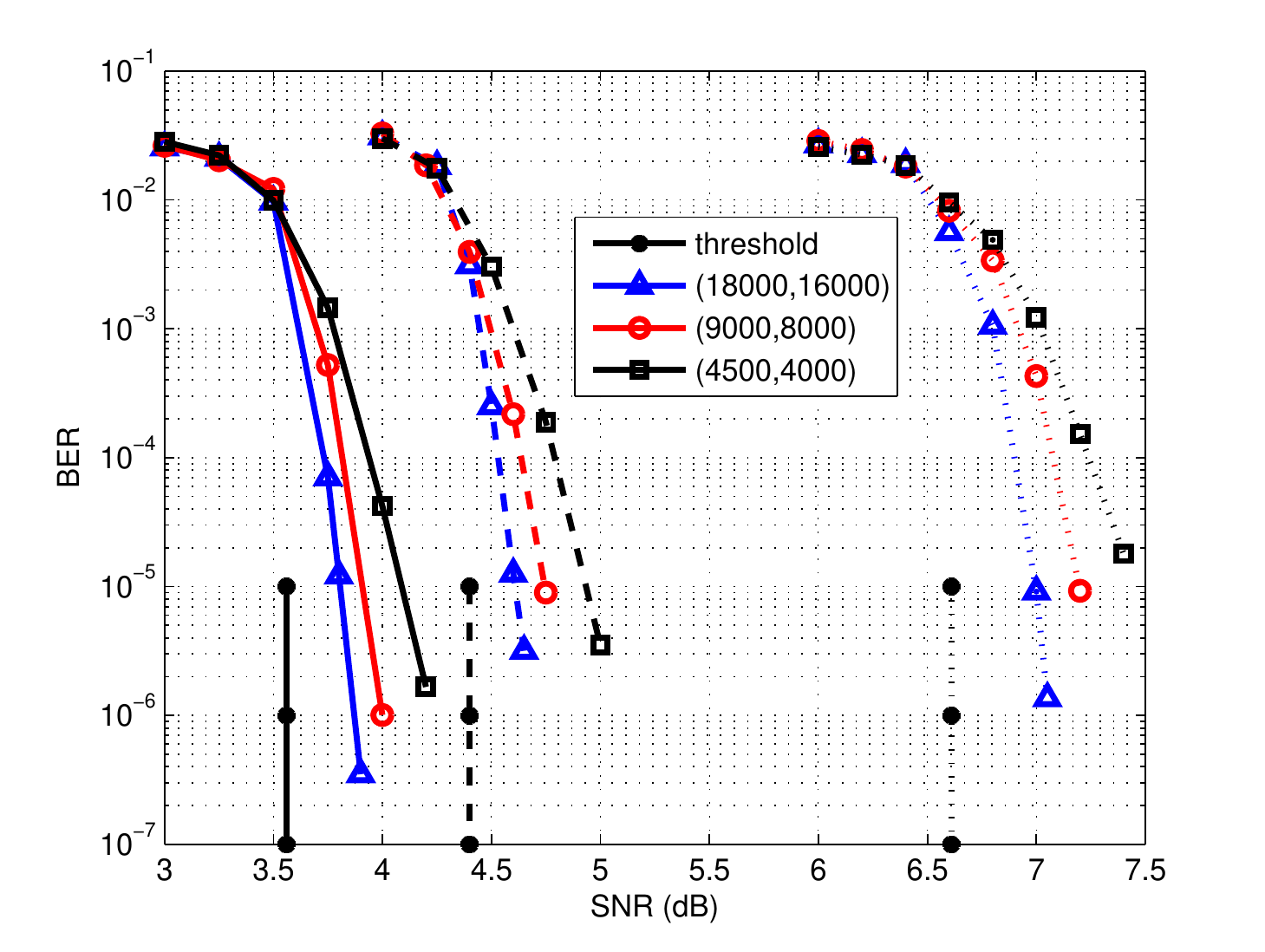}
% where an .eps filename suffix will be assumed under latex,
% and a .pdf suffix will be assumed for pdflatex; or what has been declared
% via \DeclareGraphicsExtensions.
\caption{Simulation results and thresholds for the LDPC codes with $d_v=4$, $d_c=36$ and $R=0.89$. The solid curves correspond to the LDPC codes over AWGN channel. The dashed curves correspond to the LDPC coded 2D interference system detected by JIDDS with turbo equalization. The dotted curves correspond to the LDPC coded 2D interference system detected by SSWA.}
\label{figure9}
\end{figure}

Next, we examine how tight the threshold upper bounds the performance of the system. Since the results are similar for both $\textbf{H}_A$ and $\textbf{H}_B$, we only adopt the channel $\textbf{H}_A$ in the following. Note that here the thresholds correspond to the infimum of SNR such that the message error rate converges to zero as the number of iterations tends to infinity. For a fixed 2D channel response matrix, the threshold corresponding to $\sigma$ and the threshold corresponding to SNR are one-to-one mapping. The LDPC codes adopted in the simulation are constructed using the semi-random algorithm \cite{ref4}. We denote the LDPC codes as $(N,K)$. There are two sets of LDPC codes used in our simulation. The first set contains LDPC codes $(14000,7000)$, $(6000,3000)$ and $(2000,1000)$, which are all regular LDPC codes with code rate $R=0.5$, variable node degree $d_v=3$ and check node degree $d_c=6$. The second set contains LDPC codes $(18000,16000)$, $(9000,8000)$ and $(4500,4000)$, which are all regular LDPC codes with code rate $R=0.89$, variable node degree $d_v=4$ and check node degree $d_c=36$. For the simulation, we let the coset-defining vector $\textbf{b}=\textbf{0}$. Figure \ref{figure7} shows the simulation results and the thresholds of turbo-equalized and non-turbo equalized 2D interference system detected by JIDDS. The simulation results are based on two LDPC codes, namely, $(14000,7000)$ and $(18000,16000)$. We denote the number of iterations as $I_c/I_{out}$ in the figure. $I_{out}=1$ corresponds to non-turbo equalized system and $I_{out}>1$ corresponds to the system with turbo-equalization. It can be observed that as the number of iterations increases, the simulation results approach closer to the corresponding thresholds. The effect of iterations is more significant for code $(14000,7000)$ than for code $(18000,16000)$. Figures \ref{figure8} and \ref{figure9} show the simulation results and the corresponding thresholds for the first and second set of LDPC codes, respectively. We compare the results of LDPC codes over AWGN channel, and the LDPC coded 2D interference channel detected by JIDDS and SSWA. For the AWGN channel, the SNR is defined as
\begin{eqnarray}
\label{46}
\textup{SNR}&=&10\cdot \log_{10}\left ( \frac{E_b}{N_0} \right )\nonumber\\
&=&10 \log_{10}\left ( \frac{1}{2 R \sigma ^{2}} \right ) \textup{dB}.
\end{eqnarray}
In these simulation results, the number of iterations inside the LDPC decoder is set to $I_c=50$ and the number of outer iterations between the channel detector and LDPC decoder is fixed at $I_{out}=10$. In Fig. \ref{figure8}, for the three system setups, the simulation results with code $(14000,7000)$ are about 0.4 dB away from the corresponding threshold. The simulation results and thresholds of JIDDS are about 0.8 dB worse than those of LDPC code over AWGN channel and are about 0.7 dB better than those of SSWA. In Fig. \ref{figure9}, the performance gaps between the thresholds and simulation results with code $(18000,16000)$ are less than 0.4 dB for all three system setups. This time, as the code rate increases, the simulation results and thresholds of JIDDS are about 0.9 dB worse than those of LDPC code over an AWGN channel and are about 2.2 dB better than those of SSWA. It can be observed in these two figures that as the code length $N$ increases, the simulation results approach the thresholds. When code length and number of iterations increase, the simulation result is expected to approach even closer to the threshold. The small gaps between the simulation results and the corresponding thresholds confirm the effectiveness of the density evolution technique in analyzing the performance of the system for large code lengths.

\section{conclusion}
In this paper, we have analyzed the performance of LDPC codes over 2D interference channels. The JIDDS is adopted in the 2D interference system. We have shown the graph representation of the message-flow neighborhood of JIDDS and proved that the message-flow neighborhood will be tree-like for a long code length. We have also shown two concentration theorems. The two concentration theorems state that for a particular transmitted codeword or i.u.d. random codeword, the message error probability of the system concentrates around the probability in which the message-flow neighborhood is tree-like. For the tree-like message-flow neighborhood, a modified density evolution algorithm was employed to track the message densities during the iterations. We have also used the density evolution algorithm to calculate the threshold values of the JIDDS and SSWA. For a fixed channel response matrix, the threshold represents the largest noise level the system can tolerate in order to achieve reliable transmission. Thus, whenever the standard deviation of noise is below the threshold, transmission will be reliable given that the code length and number of iterations are large enough. The simulation results have been compared to the corresponding thresholds and the gaps between them are relatively small. Therefore, the threshold calculated using density evolution can be used to accurately predict the system performance for large code lengths.

\begin{table}
\renewcommand{\arraystretch}{1.5}
\addtolength{\tabcolsep}{-5.3pt}
\centering
\caption{Evolution of $p^t_{\textup{i.u.d.}}$ during the iterations for various parameters.}
\label{Table1}
\begin{tabular}{|c|c|c|c|c|}
\hline
&\multicolumn{2}{c|}{$p^t_{\textup{i.u.d.}}$ for $\textbf{H}_A$}&\multicolumn{2}{c|}{$p^t_{\textup{i.u.d.}}$ for $\textbf{H}_B$}\\
\hline
$~~t~~$ &(3,6),$\sigma=0.81$& (4,36),$\sigma=0.45$ & (3,6),$\sigma=0.84$ & (4,36),$\sigma=0.47$\\
\hline
1 & 0.120135 & 0.024950 & 0.113098 & 0.022694\\
\hline
2 & 0.101857 & 0.019691 & 0.097696 & 0.019235\\
\hline
3 & 0.086699 & 0.013585 & 0.088068 & 0.015846\\
\hline
4 & 0.011226 & 0.000000 & 0.063708 & 0.013504\\
\hline
5 & 0.000000 & 0.000000 & 0.000000 & 0.000000\\
\hline
\end{tabular}
\end{table}
%%%%%%%%%%%%%%%%%%%%%%%%%%%%%%%%%%%%%%%%%%%%%%%%%
%%%%%%%%%%%%%%%%%%%%%%%%%%%%%%%%%%%%%%%%%%%%%%%%%
\begin{table}
\renewcommand{\arraystretch}{1.5}
\addtolength{\tabcolsep}{-5.3pt}
\centering
\caption{Threshold values for regular LDPC codes over AWGN channel and 2D ISI channel.}
\label{Table2}
\begin{tabular}{|c|c|c|c|c|c|c|c|c|}
\hline
&&AWGN&\multicolumn{3}{c|}{2D ISI: $\textbf{H}_A$}&\multicolumn{3}{c|}{2D ISI: $\textbf{H}_B$}\\
\hline
$(d_v,d_c)$ & ~$R$~ & $\sigma_{AWGN}$ & ~~$\sigma_{te}$~~ & $\sigma_{non-te}$& $\sigma_{SSWA}$&~~$\sigma_{te}$~~& $\sigma_{non-te}$& $\sigma_{SSWA}$\\
\hline
$(3,4)$ & 0.25 & 1.26 & 1.19 & 1.10 & 1.12 &1.22 &1.15 &1.14\\
\hline
$(3,6)$ & 0.50 & 0.88 & 0.81& 0.73 & 0.74&0.84 &0.78 &0.75\\
\hline
$(3,10)$ & 0.70 & 0.68 & 0.62& 0.56 &0.54& 0.64&0.59&0.53\\
\hline
$(3,20)$ & 0.85 & 0.54 & 0.50& 0.46 &0.39& 0.52&0.49 &0.38\\
\hline
$(4,36)$ & 0.89 & 0.50 & 0.45& 0.43 &0.35& 0.47& 0.46&0.33\\
\hline
\end{tabular}
\end{table}
%%%%%%%%%%%%%%%%%%%%%%%%%%%%%%%%%%%%%%%%%%%%%%%%%

\begin{thebibliography}{1}
\bibitem{ref1} C. Berrou, A. Glavieux, and P. Thitimajshima, ``Near Shannon limit error-correcting coding and decoding," in \emph{Proc. IEEE Int. Conf. on Communications (ICC)}, May 1993, pp. 1064-1070.
\bibitem{ref2} R. G. Gallager, \emph{Low-Density Parity-Check Codes}. Cambridge, MA: MIT Press, 1963.
\bibitem{ref3} D. J. C. MacKay and R. M. Neal,``Near Shannon limit performance of low-density parity-check codes," \emph{Electron. Lett.}, vol. 32, no. 18, pp. 1645-1646, Aug. 1996.
\bibitem{ref4} D. J. C. MacKay, ``Good error correcting codes based on very sparse matrices," \emph{IEEE Trans. Inform. Theory}, vol. 45, no. 2, pp. 399-431, Mar. 1999.
\bibitem{ref5} N. Alon and M. Luby, ``A linear time erasure-resilient code with nearly optimal recovery," \emph{IEEE Trans. Inform. Theory}, vol. 42, no. 6, pp. 1732-1736, Nov. 1996.
\bibitem{ref6} M. G. Luby, M. Mitzenmacher, M. A. Shokrollahi, and D. A. Spielman, ``Improved low-density parity-check codes using irregular graphs and belief propagation," in \emph{Proc. IEEE Int. Symp. Information Theory (ISIT)}, Cambridge, MA, USA, Aug. 1998, pp. 117
\bibitem{ref7} M. G. Luby, M. Mitzenmacher, M. A. Shokrollahi, and D. A. Spielman, ``Improved low-density parity-check codes using irregular graphs," \emph{IEEE Trans. Inform. Theory}, vol. 47, no. 2, pp. 585-598, Feb. 2001.
\bibitem{ref8} T. J. Richardson and R. L. Urbanke, ``The capacity of low-density parity-check codes under message-passing decoding," \emph{IEEE Trans. Inform. Theory}, vol. 47, no. 2, pp. 599-618, Feb. 2001.
\bibitem{ref9} S. -Y. Chung, G. D. Forney, Jr., T. J. Richardson, and R. Urbanke, ``On the design of low-density parity-check codes within 0.0045 dB of the Shannon limit," \emph{IEEE Commun. Lett.}, vol. 5, no. 2, pp. 58-60, Feb. 2001.
\bibitem{ref10} S. -Y. Chung, T. J. Richardson, and R. L. Urbanke, ``Analysis of sum-product decoding of low-density parity-check codes using a Gaussian approximation," \emph{IEEE Trans. Inform. Theory}, vol. 47, no. 2, pp. 657-670, Feb. 2001.
\bibitem{ref11} T. Mittelholzer, A. Dholakia, and E. Eleftheriou, ``Reduced-complexity decoding of low density parity check codes for generalized partial response channels," \emph{IEEE Trans. Magn.}, vol. 37, no. 2, pp. 721-728, Mar. 2001.
\bibitem{ref12} B. M. Kurkoski, P. H. Siegel, and J. K. Wolf, ``Joint message-passing decoding of LDPC codes and partial-response channels," \emph{IEEE Trans. Inform. Theory}, vol. 48, no. 6, pp. 1410-1422, June 2002.
\bibitem{ref13} H. Song and B. V. K. V. Kumar, ``Low-density parity-check codes for partial-response channels," \emph{IEEE Signal Process. Mag.}, vol. 21, no. 1, pp. 56-66, Jan. 2004.
\bibitem{ref14} A. Kavcic, X. Ma, and M. Mitzenmacher, ``Binary intersymbol interference channels: Gallager codes, density evolution, and code performance bounds," \emph{IEEE Trans. Inform. Theory}, vol. 49, no. 7, pp. 1636-1652, July 2003.
\bibitem{ref15} W. Tan and J. R. Cruz, ``Performance evaluation of low-density parity-check codes on partial-response channels using density evolution," \emph{IEEE Trans. Comm.}, vol. 52, no. 8, pp. 1253-1256, Aug. 2004.
\bibitem{ref16} A. Thangaraj and S. W. McLaughlin, ``Thresholds and scheduling for LDPC-coded partial response channels," \emph{IEEE Trans. Magn.}, vol. 38, no. 5, pp. 2307-2309, Sep. 2002.
\bibitem{ref17} S. J. Greaves, Y. Kanai, and H. Muraoka, ``Magnetic recording in patterned media at 5-10 $Tb/in^2$," \emph{IEEE Trans. Magn.}, vol. 44, no. 11, pp. 3430¨C3433, 2008.
\bibitem{YaoWang2D} Y. Wang, J. Yao, and B. V. K. V. Kumar, ``2-D write/read channel model for bit-patterned media recording with large media noise," \emph{IEEE Trans. Magn.}, vol. 51, no. 12, pp. 1-11, 2015.
\bibitem{ref18} R. Wood, M. Williams, A. Kavcic, and J. Miles, ``The feasibility of magnetic recording at 10 Terabits per square inch on conventional media," \emph{IEEE Trans. Magn.}, vol. 45, no. 2, pp. 917-923, Feb. 2009.
\bibitem{JYAOTDMR} J. Yao, E. Hwang, B. V. K. V. Kumar, and G. Mathew, ``Two-track joint detection for two-dimensional magnetic recording (TDMR)," in \emph{Proc. IEEE Int. Conf. Commun., 2015}. (ICC'15), pp. 418-424.
\bibitem{ref19} M. H. Kryder, E. C. Gage, T. W. McDaniel, W. A. Challener, R. E. Rottmayer, G. Ju, Y.-T. Hsia, and M. F. Erden, ``Heat assisted magnetic recording," \emph{Proc. of the IEEE}, vol. 96, no. 11, pp. 1810-1835, 2008.
%\bibitem{ref20} H. J. Coufal, G. T. Sincerbox, and D. Psaltis, Eds., \emph{Holographic Data Storage.} Secaucus, NJ: Springer- Verlag, 2000.
\bibitem{viterbi} A. J. Viterbi, ``Error bounds for convolutional codes and an asymptotically optimum decoding algorithm," \emph{IEEE Trans. Inform. Theory}, vol. 13, pp. 260-269, Apr. 1967.
\bibitem{BCJR} L. R. Bahl, J. Cocke, F. Jelinek, and J. Raviv, ``Optimal decoding of linear codes for minimizing symbol error rate," \emph{IEEE Trans. Inform. Theory}, vol. 20, pp. 284-287, Mar. 1974.
\bibitem{NP_complete} E. Ordentlich and R. M. Roth, ``On the computational complexity of 2D maximum-likelihood sequence detection," HP Labs, 2006, Tech. Rep. HPL-2006-69.
\bibitem{JYAO} J. Yao, K. C. Teh, and K. H. Li, ``Performance evaluation of maximum-likelihood page detection for 2D interference channel," \emph{IEEE Trans. Magn.}, vol. 48, no. 7, pp. 2239-2242, July 2012.
\bibitem{NSingla} N. Singla and J. A. O' Sullivan, ``Joint equalization and decoding for nonlinear two-dimensional intersymbol interference channels," in \emph{Proc. IEEE Int. Symp. Information Theory (ISIT)}, Adelaide, Australia, Sep. 2005, pp. 1353-1357.
\bibitem{ref21} S. Nabavi and B. V. K. V. Kumar, ``Two-dimensional generalized partial response equalizer for bit-patterned media," in \emph{Proc. IEEE Int. Conf. Commun. (ICC)}, June 2007, pp. 6249-6254.
\bibitem{ref22} W. Chang and J. R. Cruz, ``Inter-track interference mitigation for bit-patterned magnetic recording," \emph{IEEE Trans. Magn.}, vol. 46, no. 11, pp. 3899-3908, Nov. 2010.
\bibitem{ref23} S. Karakulak, P. H. Siegel, J. K. Wolf, and H. N. Bertram, ``Joint-track equalization and detection for bit patterned media recording," \emph{IEEE Trans. Magn.}, vol. 46, no. 9, pp. 3639-3647, Sep. 2010.
\bibitem{JYaoBCJR} J. Yao, K. C. Teh, and K. H. Li, ``Reduced-state Bahl--Cocke--Jalinek--Raviv detector for patterned media storage," \emph{IEEE Trans. Magn.}, vol. 46, no. 12, pp. 4108-4110, 2010.
\bibitem{ref24} X. Chen and K. M. Chugg, ``Near-optimal data detection for two-dimensional ISI/AWGN channels using concatenated modeling and iterative algorithms," in \emph{Proc. IEEE Int. Conf. Commun. (ICC)}, June 1998, pp. 952-956.
\bibitem{ref25} Y. Wu, J. A. O'Sullivan, N. Singla, and R. S. Indeck, ``Iterative detection and decoding for separable two-dimensional intersymbol interference," \emph{IEEE Trans. Magn.}, vol. 39, no. 4, pp. 2115-2120, July 2003.
\bibitem{IRCSDF1} T. Cheng, B. J. Belzer, and K. Sivakumar, ``Row-column soft-decision feedback algorithm for two-dimensional intersymbol interference," \emph{IEEE Signal Process. Lett.}, vol. 14, no. 7, pp. 433-436, July 2007.
\bibitem{IRCSDF2} Y. Zhu, T. Cheng, K. Sivakumar, and B. J. Belzer, ``Markov random field detection on two-dimensional intersymbol interference channels," \emph{IEEE Trans. Signal Process.}, vol. 56, no. 7, pp. 2639-2648, July 2008.
\bibitem{MCMC} M. Shaghaghi, K. Cai, Y. L. Guan, and Z. L. Qin, ``Markov chain Monte Carlo based detection for two-dimensional intersymbol interference channels," \emph{IEEE Trans. Magn.}, vol. 47, no. 2, pp. 471-478, Feb. 2011.
\bibitem{JYAO2} J. Yao, K. C. Teh, and K. H. Li, ``Joint iterative detection/decoding scheme for two-dimensional interference channels," \emph{IEEE Trans. Commun.}., vol. 60, no. 12, pp. 3548-3555, 2012.
\bibitem{JYAO3} J. Yao, K. C. Teh, and K. H. Li, ``Joint message-passing decoding of LDPC codes and 2-D ISI channels,¡± \emph{IEEE Trans. Magn.}, vol. 49, no. 2, pp. 675-681, 2013.
\bibitem{JYAO4} J. Yao, K. C. Teh, and K. H. Li, ``Iterative detection scheme with LDPC codes for two-dimensional interference channels,¡± \emph{2012 IEEE International Conference on Commun. Systems} (ICCS), IEEE, 2012, pp. 398-402.
\bibitem{ref26} P. Robertson, E. Villebrun, and P. Hoeher, ``A comparison of optimal and suboptimal MAP decoding algorithms operating in the log domain," \emph{Proc. Int. Conf. on Comm.}, 1995, pp. 1009-1013.
\bibitem{ref27} X-Y. Hu, E. Eleftherious, D-M. Arnold, and A. Dholakia, ``Efficient implementation of the sum-product algorithm for decoding LDPC codes," \emph{Proc. 2001 IEEE GlobeCom Conf.}, Nov. 2001, pp. 1036-1036E.
\end{thebibliography}
\end{document}